\documentstyle[12pt]{article}
\begin{document} 
\thispagestyle{empty} 

\begin{center}
{\large{\bf DERIVING THE QGP-HADRON TRANSITION CURVE
FROM TWO SEPARATE PARTITION FUNCTIONS\\}}
\vspace{2cm} 
{\large A. S. Kapoyannis}\\ 
\smallskip 
{\it Department of Physics, University of Athens, 15771 Athens, Greece}\\
\vspace{1cm}

\end{center}
\vspace{0.5cm}
\begin{abstract}
A method is developed to consistently satisfy the Gibbs equilibrium
conditions between the quark-gluon and hadronic phase although each
phase has been formulated in separate grand canonical partition
function containing three quark flavours. The sector in the space of
thermodynamic variables where the transition takes place is restricted to
a curve, according to the phase diagram of QCD. The conservation laws of
quantum numbers are also imposed on the transition curve. The effect of the
inclusion of the newly discovered pentaquark states is considered.
The freeze-out conditions of $S+S$, $S+Ag$ (SPS) and $Au+Au$ (RHIC) are found
compatible with a primordial QGP phase, but the conditions indicated by
$Pb+Pb$ (SPS) are not.
\end{abstract}

\vspace{3cm}



{\it e-mail address:} akapog@cc.uoa.gr (A.S. Kapoyannis)

\newpage
\setcounter{page}{1}

\vspace{0.3cm}
{\large \bf 1. Introduction}

Quantum Chromodynamics is universally accepted as the theory 
of strong interactions. Within the context of this
theory the phase of Quark-Gluon plasma receives accurate description.
However, the formation of the hadronic phase, which is the final state of
any possible primordial QGP state, still remains an open
problem in view of QCD. On the other hand the hadronic multiplicities
emerging from heavy-ion collisions have been extensively and successfully
predicted by statistical models using a handful of thermodynamical
parameters [1-8]. So the use of two separate models for the QGP and the hadron
phase, called Hadron Gas (HG), offers a complementary approach.

QCD predicts that the transition between QGP and hadronic phase is of first
order at high baryon densities (depicted by a critical line
on the ($T,\mu_B$) plane), while it is of higher order at small or zero
baryon densities (crossover). The end point of the first order transition line is a
critical point [9]. The transition  points must be restricted to a curve on
the phase diagram of temperature and baryon chemical potential. In view of
this aspect any models to be used for the
description of QGP and HG have to be matched properly at the transition
between the two phases.

The aim of this work is to trace the sector of the space of thermodynamic
variables where the QGP-hadron transition occurs, with the following
requirements:
a) Any mixed phase formed in the first order part
of the transition must occupy only a curve in the space of the thermodynamic
variables. This requirement is even more strong in the crossover area where
a mixed phase does not exist.
b) The Gibbs equilibrium conditions have to be satisfied, which amount to
$T_{QGP}=T_{HG}$ for thermal equilibrium, $P_{QGP}=P_{HG}$ for mechanical
equilibrium and
$\{\mu\}_{QGP}=\{\mu\}_{HG}$ for chemical equilibrium, where $\{\mu\}$
stands for the set of chemical potentials used in the description of the
two phases.
c) All the conservation laws of quantum numbers like baryon number $B$, electric
charge $Q$, strangeness $S$, etc. have to be satisfied at every point on the
transition line, in a way that they could be extended for every number of
flavours that are present to the system.

These problems are confronted every time separate partition functions are
used for the two phases, but the simultaneous fulfilment of the above
conditions has not been achieved.
Among the numerous examples that exist, in [10], where
only light, identical quarks are used ($u=d \equiv q$), the curve of equal pressures are
made to approximately coincide by a choice on the external parameters $B$ and
$a_s$, something which does not allow matching when other flavours are
introduced. In [11] the strange fugacity $\lambda_s$ is discontinuous at the
HG-QGP transition and the conservation of baryon number can only be
accommodated at the case of first order transition.
In [12] the strange chemical potential $\mu_s$ is also
discontinuous. In [13] only $q$ quarks are considered and the
requirement of continuity of chemical potentials and conservation of baryon
number leads to a mixed phase which occupies a surface and not a line on
the ($T,\mu_B$) plane.
The same is true in [14-16] where also $s$ quarks are included. 
In [17] there is an analogous situation as in [13]
with a critical line at the ($T,\mu_B$) plane
but the conservation of baryon number is not considered.
In [18] the $q$ and $s$ quark chemical potentials are continuous but baryon
number and strangeness of the system are not kept constant during
hadronisation, since hadrons evaporate from QGP. The considerations of [11-18]
are consistent with a first order transition but cannot be valid at the
crossover region.
In this work all the thermodynamic variables and the pressure will be kept
continuous at the transition line (in contrast with [10-12]), the first order
part of the transition will be presented by a line on the ($T,\mu_B$) plane
(differing from [13-16]), in the mixed phase the quantum numbers will be
conserved to each constituent phase (differing from [14]) and no evaporation
of hadrons will be assumed from the system (differing from [18]).

Let us consider the requirements that a system with $n_f$ quark
flavours has to satisfy. Every
conservation law accounts for two equations to be fulfilled. One
sets the value of the quantum quantity, e.g. $<\!B\!>_{QGP}=b_i$ and the other
assures the conservation at the phase transition, e.g. $<\!B\!>_{QGP}=<\!B\!>_{HG}$.
The total number of equations that must hold are, thus, $2n_f+1$ (the unit
accounts for the equality of pressures). Assuming full chemical equilibrium
every quark flavour introduces one extra fugacity in the set of the
thermodynamical variables, which, with the inclusion of volume and
temperature, amount to $n_f+2$. At the crossover region, the surviving free
parameters to fulfil the necessary equations decrease to $n_f+1$,
since the equality of volume $V_{QGP}=V_{HG}$ results to
the equality of densities between the two phases. At the first order
transition line the free parameters are $n_f+2$, since now
$V_{QGP} \neq V_{HG}$. It is evident then that the necessary $2n_f+1$
conditions can be fulfilled only at the first-order part of the transition
and only when there is one flavour present, $n_f=1$, or when the $u$ and $d$
quarks are considered identical ($q$ quarks, described by a single chemical
potential $\mu_q$). It has to be clarified that the conditions like
$<B>_{QGP}=b_i$ have to be satisfied in order to have a whole line of
transition points. If these equations are dropped then we are left with
$n_f+1$ equations, which can be solved, but result to a unique point in the
space of the thermodynamical parameters.

\vspace{0.3cm}
{\large \bf 2. Expanding the fugacity sector}

It is clear that in order to satisfy $2n_f+1$ relations, every
flavour has to be accompanied by two fugacities instead of one.
The multiplicity data emerging from heavy ion collisions suggest that the
thermalised hadronic system has not achieved full chemical equilibrium.
First the strangeness partial chemical equilibrium factor $\gamma_s$ had been
introduced [2] and used extensively to model the data [3-4]. Also a similar
factor for the light quarks $\gamma_q$ was introduced [5] and used in many
analyses [6]. Here the light $u$ and $d$ quarks will be
accompanied by separate fugacities $\gamma_u$, $\gamma_d$. A factor
$\gamma_j$ controls the quark density $n_j+n_{\bar{j}}$ in contrast to the
usual fugacity $\lambda_j$ which controls the net quark density
$n_j-n_{\bar{j}}$ [3]. These additional fugacities can serve the
purpose of satisfying the necessary equations at the transition point, as well
as, preserving the continuity of chemical potentials between the two phases.

A system with 3 flavours ($u$, $d$ and $s$ quarks)
is described by the set of thermodynamical variables
$(T,\lambda_u,\gamma_u,\lambda_d,\gamma_d,\lambda_s,\gamma_s)\equiv$
$(T,\{\lambda,\gamma\})$.
Assuming strangeness neutrality at the QGP phase leads to $\lambda_s=1$.
Setting $x=V_{HG}/V_{QGP}$, the set of
equations to be satisfied at every phase transition point will be
\begin{equation}
P_{QGP}(T,\{\lambda,\gamma\}_{\lambda_s=1})=
P_{HG}(T,\{\lambda,\gamma\}_{\lambda_s=1})
\end{equation}
\begin{equation}
<\!n_B\!>_{QGP}(T,\{\lambda,\gamma\}_{\lambda_s=1})=
x<\!n_B\!>_{HG}(T,\{\lambda,\gamma\}_{\lambda_s=1})
\end{equation}
\begin{equation}
<\!n_B\!>_{QGP}(T,\{\lambda,\gamma\}_{\lambda_s=1})=
2\beta<\!n_Q\!>_{QGP}(T,\{\lambda,\gamma\}_{\lambda_s=1})
\end{equation}
\begin{equation}
<\!n_Q\!>_{QGP}(T,\{\lambda,\gamma\}_{\lambda_s=1})=
x<\!n_Q\!>_{HG}(T,\{\lambda,\gamma\}_{\lambda_s=1})
\end{equation}
\begin{equation}
<\!n_S\!>_{HG}(T,\{\lambda,\gamma\}_{\lambda_s=1})=0\;,
\end{equation}
where $n$ denotes densities.
For isospin symmetric systems one has to set $\beta=1$ in (3). Eqs. (1)-(5)
have only one free variable, necessary to produce a whole transition line in
the phase diagram. At crossover $x=1$, whereas at the first order transition
line the inequality $V_{QGP} \neq V_{HG}$ preserves the survival of $x$ as an
extra variable.

\vspace{0.3cm}
{\large \bf 3. A solution for the transition curve}

Two simple models, will now be employed to apply the above considerations,
the whole approach is, though, applicable to every partition function used.
For the Hadron Gas phase all the hadronic states containing $u$, $d$ and $s$
quarks will be used [19]. The Bose-Einstein or Fermi-Dirac statistics
applicable to each hadron will also be taken into account.
Thus the HG partition function for point particles can be written down as
\begin{equation}
\ln Z_{HG\;pt}(V,T,\{\lambda,\gamma\})=
\frac{V}{6\pi^2 T}
\sum_{\rm a} \sum_i g_{{\rm a}i}
\int_0^{\infty} \frac{p^4}{\sqrt{p^2+m_{{\rm a}i}^2}}
\frac{1}{e^{\sqrt{p^2+m_{{\rm a}i}^2}/T}\lambda_{\rm a}^{-1}+\alpha}dp\;,
\end{equation}
where $g_{{\rm a}i}$ are degeneracy factors due to spin and isospin and
$\alpha=-1(1)$ for bosons (fermions).
The index ${\rm a}$ runs over all hadronic families, each of which contains
members with the same quark content and $i$ to all the particles of this
family. The fugacity
$\lambda_{\rm a}=\prod_j\lambda_j^{N_j-N_{\bar{j}}} \gamma_j^{N_j+N_{\bar{j}}}$,
where $j=u,d,s$ and $N_j(N_{\bar{j}})$ is the number of $j(\bar{j})$ quarks
contained in a hadron belonging to family ${\rm a}$. 

The repulsion which is due to the finite hadron size can be taken into
account.
Assuming that each hadrons' volume is proportional to its mass,
$V_{{\rm a}i}/m_{{\rm a}i}=V_0$ the following factor can be calculated
\begin{equation}
f=\sum_{\rm a} \sum_i n_{{\rm a}i} V_{{\rm a}i}=
\frac{V_0}{6\pi^2 T}
\sum_{\rm a} \sum_i g_{{\rm a}i}m_{{\rm a}i}
\int_0^{\infty} \frac{p^4}{\sqrt{p^2+m_{{\rm a}i}^2}}
\frac{1}{e^{\sqrt{p^2+m_{{\rm a}i}^2}/T}\lambda_{\rm a}^{-1}+\alpha}dp\;,
\end{equation}
with $V_0$ remaining an open parameter controlling the hadron size. The
pressure and the densities are then evaluated by
\begin{equation}
P_{HG}=\frac{P_{HG\;pt}}{1+f}\hspace{3cm}
n_{i\;HG}=\frac{n_{i\;HG\;pt}}{1+f}
\end{equation}

For the QGP phase a simple model containing 3 flavours is used. The quarks
are non-interacting and only the presence of gluons is accounted for, as well
as the effect of the vacuum through the MIT bag constant, $B$. A wealth of quark
fugacities is easily accommodated, though, in this model. The QGP partition
function is consequently
\begin{equation}
\ln Z_{QGP}(V,T,\{\lambda,\gamma\})=
\frac{N_s N_c V}{6\pi^2 T}
\sum_j \int_0^{\infty} \frac{p^4}{\sqrt{p^2+m_j^2}}
\frac{1}{e^{\sqrt{p^2+m_j^2}/T}(\lambda_j \gamma_j)^{-1}+1} dp
+V\frac{8\pi^2 T^3}{45}-\frac{BV}{T}\;,
\end{equation}
where $N_s=2$ and $N_c=3$.
The index $j$ runs to all quarks and antiquarks and the fugacity 
$\lambda_{\bar{j}}=\lambda_j^{-1}$ and $\gamma_{\bar{j}}=\gamma_j$. The
current quark masses are $m_u=1.5$, $m_d=6.75$ and $m_s=117.5$ MeV [19].

Eqs.~(1)-(5) are then solved for $x=1$ for the crossover region.
At the first-order QGP-HG transition a mixed phase is assumed
\begin{equation}
\ln Z_{mixed}(\delta,T,\{\lambda,\gamma\})=
\delta \ln Z_{QGP}(V_{QGP},T,\{\lambda,\gamma\})+
(1-\delta) \ln Z_{HG}(V_{HG},T,\{\lambda,\gamma\})\;,
\end{equation}
where the set of fugacities $\{\lambda,\gamma\}$ are
kept constant through out the first-order transition line, although the
densities are not. The parameter
$0 \le \delta \le 1$ and for $\delta=0(\delta=1)$ we have pure HG(QGP) phase.
Eqs.~(1)-(5) are solved for the pure phases leading to the
determination of the thermodynamic parameters. The constraints are then
automatically satisfied for every point in the mixed phase. For example
eq.~(2) is equivalent to
$<\!B\!>_{QGP}(V_{QGP},T,\{\lambda,\gamma\})$$=<\!B\!>_{HG}(V_{HG},T,\{\lambda,\gamma\})$.
Thus $<\!B\!>_{mixed}$$=\delta<\!B\!>_{QGP}+(1-\delta)<\!B\!>_{HG}$$=<\!B\!>_{QGP}$$=<\!B\!>_{HG}$.
The pressure of the mixed phase is also kept constant. Through the validity
of eq.~(1) $P_{mixed}$$=\delta P_{QGP}+(1-\delta) P_{HG}$$=P_{QGP}$$=P_{HG}$.

The HG partition function (6)-(8) and the QGP partition function (9) is
used to the system of eqs.~(1)-(5). This system accepts as solution for the
variable $\gamma_s$ the value 0, since then eq.~(5) is automatically satisfied.
This a trivial solution because it is equivalent to the absence of the
strange quarks in the system and so such solutions should be excluded. A
non-trivial solution for the thermodynamic variables is
depicted for the parameters $B^{1/4}=280$ MeV and $V_0=1.6/(4B)$.
The position of the critical point is set at $\mu_{B\;cr.p.}=360$ MeV as in
[20]. For the ratio of the volumes $x=V_{HG}/V_{QGP}$ the adopted form for
$\mu_B>\mu_{B\;cr.p.}$ is
$x=1+\left(\frac{\mu_B-\mu_{B\;cr.p.}}{1000-\mu_{B\;cr.p.}}\right)^2 0.05$.
It should be pointed
that the position of the critical point and the expansion ratio in the 1st
order transition cannot be predicted by the simple model used for these
calculations, due to the absence of interaction from both phases. The
temperature $T$ is displayed as function of the baryon chemical potential
$\mu_B$ in Fig.~1. The critical point is also depicted and it divides the
crossover line (slashed line) from the first order transition line (solid
line). The relative chemical equilibrium fugacity $\gamma_u$ is displayed as
function of $\mu_B$ in Fig.~2. This particular
solution leads to the gradual suppression of $\gamma_u$ as baryon chemical
potential increases. The connection of $\gamma_u$ and $\gamma_d$ for isospin
symmetric solution to both QGP and HG phases is depicted in Fig.~3. The line
$\gamma_u=\gamma_d$ is also drawn for comparison.
The relative chemical
equilibrium factor $\gamma_s$ is drawn as function of baryon chemical
potential in Fig.~4. The solution has a part with suppressed $\gamma_s$ and
another with enhanced $\gamma_s$, but the values are not far from unit.

One direct consequence of the simultaneous solution of eqs.~(1)-(5) is that
the relative chemical equilibrium fugacities have values that depend on each
other at every transition point. This is easily realised in the condition
$<\!S\!>_{HG}=0$ (for $\lambda_s=1$). The solution of this condition is greatly
simplified by the use of the Boltzmann approximation and the assumption that
isospin symmetry leads to the approximate solution
$\lambda_u=\lambda_d\equiv\lambda_q$ and $\gamma_u=\gamma_d\equiv\gamma_q$.
Neglecting trivial solutions, where $\gamma_s = 0$, the zero strangeness
condition can be solved to give
\begin{equation}
\gamma_s=
\frac{F_K(T)-F_H(T)\gamma_q(\lambda_q+\lambda_q^{-1})}{2F_{\Xi}(T)}\;.
\end{equation}

In eq.~(11) $F_K$ represents the Kaon mesons, $F_H$  the Hyperon baryons
($\Lambda$'s and $\Sigma$'s) and $F_{\Xi}$ the $\Xi$ baryons and the
summation
\begin{equation}
F_{\rm a}(T)=\frac{T}{2\pi^2}
\sum_i g_{{\rm a}i} m_{{\rm a}i}^2 K_2(\frac{m_{{\rm a}i}}{T})
\end{equation}
includes all members of every family.
In the above relation $K$ denotes a modified Bessel function of the second
kind. It is evident from eq.~(11) that the increase of the relative chemical
equilibrium factor for light quarks, $\gamma_q$ and the increase of the light
quark fugacity $\lambda_q$ leads, at constant temperature, to the decrease of
the factor $\gamma_s$ and, thus, to the strange content of the system
at the transition point.

\vspace{0.3cm}
{\large \bf 4. Inclusion of pentaquarks}

Recently there has been evidence of hadrons containing five quarks. These
5-quark states are the $\Theta^+(1540)$ [21] with
$I=0$ and quark content $uudd\bar{s}$ and the $\Xi^*(1862)$ with $I=3/2$.
The content of the states $\Xi^*(1862)$ is $ssdd\bar{u}$ (for the state with
electric charge Q=-2), $ssud\bar{u}$ (with Q=-1), $ssud\bar{d}$ (with Q=0)
and $ssuu\bar{d}$ (with Q=+1). The existence of the first three of the states
$\Xi^*(1862)$ has been confirmed [22]. Due to the quark content of these
states the eqs. (1)-(5) are altered. This can easily be realised if the
corresponding equation of (11) is written down as
\begin{equation}
\gamma_s=
\frac{F_K(T)+F_{\Theta}(T)\gamma_q^3(\lambda_q^2+\lambda_q^{-2})
(\lambda_q+\lambda_q^{-1})
-F_H(T)\gamma_q(\lambda_q+\lambda_q^{-1})}
{2[F_{\Xi}(T)+F_{\Xi^*}(T)\gamma_q^2]}\;.
\end{equation}
The existence of $\Theta$ hadron drives $\gamma_s$ to higher values with a
strong dependence on $\gamma_q$ and $\lambda_q$, whereas the inclusion of the
$\Xi^*$ states contribute to decrease of $\gamma_s$.

The system of eqs. (1)-(5) is then solved with the inclusion of the
$\Theta^+(1540)$ and $\Xi^*(1862)$ states\footnote{The partition function of
eqs.~(6)-(8) has been used for the HG phase and the partition function of
eq.~(9) for the QGP phase.} for the same parameters $B$, $V_0$ and the same
volume ratio $x$ as in the case
without the inclusion of the pentaquarks. The results are shown in Figs.~1-4.
It is evident that, now, at the transition line $\gamma_s$ has increased
compared to the case when the 5-quarks were neglected. Also,
$\gamma_s>1$, which is in agreement with the enhancement of strangeness
production in the QGP phase.

\vspace{0.3cm}
{\large \bf 5. Application to heavy-ion data}

The quantum constraints previously discussed can be used as a diagnostic
tool for a primordial QGP phase. Assuming (a) that a quark-gluon phase has
been formed in a collision experiment and (b) that the chemical freeze-out
occurs right after the transition to the hadronic phase, then the freeze-out
thermodynamic variables have to fulfil eqs.~(1)-(5).
If, on the contrary, no quark-gluon state is formed before hadronization,
then, the restriction on the freeze-out conditions of the system is
diminished.
The thermodynamic variables are
extracted through a fit of the experimentally measured particle
multiplicities or ratios to a statistical model. Such a technique has
been successful. If now the additional constraints (1)-(5) are
imposed, the question that arises is whether a successful fit is also
produced or the variables that these constraints imply are inconsistent with
the data.

It is easier for the fitting procedure to form a subset of the necessary
equations and
apply them in order to determine first a subset of the available parameters.
Eq.~(4), when eqs.~(2) and (3) are valid, can equivalently be rewritten in
the form
\begin{equation}
<\!n_B\!>_{HG}(T,\{\lambda,\gamma\}_{\lambda_s=1})=
2\beta<\!n_Q\!>_{HG}(T,\{\lambda,\gamma\}_{\lambda_s=1})
\end{equation}
Eqs.~(3), (5) and (14) (referred to as Set B) now form a set of equations
that do not depend on
the parameters $V_0$ for the particle size, $B$ (MIT bag constant) nor the
ratio $x$ applicable to the first order transition line. Especially eq.~(5)
is completely model independent and under the assumptions of isospin
symmetry and Boltzmann statistics can acquire the simple form (11) or (13).
Eq.~(14) is also model independent since the volume corrections have been
cancelled out. Eq.~(3) depends only on the quark masses.

On the contrary, eqs.~(1) and (2) are model dependent and contain unknown
parameters. However, if the freeze-out parameters are determined they can be
inserted to eq.~(2) to determine $V_0$ (assuming that $x$ is known) and
then eq.~(1) can be used to determine $B$. This task serves to show that
eqs.~(1) and (2) have a real solution and contributes to the overall
consistency of the technique.

The constraints of set B have been used in the search for
the freeze out parameters with data from the experiments $S+S$ [23], $S+Ag$ [24]
(NA35) at beam energy $200$ AGeV, $Pb+Pb$ [25] (NA49) at beam energy $158$ AGeV
and $Au+Au$ [26] (STAR) at $\sqrt{s_{NN}}=130$ GeV. The data used are listed in
Table 1 and they are in all cases full phase space multiplicities except from
the RHIC data which are measured in the midrapidity. The experiments are
so chosen because they do not produce great baryon chemical potential at
freeze out and so they are probably at the crossover area [20], allowing one
to set $x=1$. The technique can be applied to the first order transition case,
determining the freeze-out variables, but then the unimportant parameters
$V_0$ and $B$ cannot not be uniquely determined.

The theoretical calculation of the particle multiplicity necessary to perform
a fit to the experimental data has been carried out with the partition
function (6), (7). The right
Bose-Einstein or Fermi-Dirac statistics for every particle has been used
throughout the calculations. The feeds from the decay of resonances have also
been included.

The results from the fits performed are listed in Table 2.
The set of constraints B includes the conditions that the system freezes on
the QGP-Hadron transition line. For
comparison, the $\chi^2/dof$ result from the fit with only the constraints
relevant to the hadronic phase (eqs.~(5) and (14) with $\lambda_s$ left as
a free parameter) is also listed in the sector of set A. The value of $\beta$
is set to 1 in the case of $S+S$, 1.1 in the case of $S+Ag$, 1.27 for
$Pb+Pb$ and 1.25 for $Au+Au$. Two fits are performed in each case, one with
all the multiplicities included and one without the multiplicity that
contains the pions. The reason is that the inclusion of this multiplicity
deteriorates the quality of the fit [8,27]\footnote{The
presence of excess of pions, though, can be connected with a primordial
high entropy phase or with a phase with the chiral symmetry restored [28].}. So a bad
fit, when the additional
constraints of set B are imposed, may be partly due to the presence of this
multiplicity. For this reason the fit without the pions is more reliable.

For the $S+S$ and $S+Ag$ data the quality of the fit with set B is of medium
quality ($\chi^2/dof=2.94$ and $1.91$, respectively) when the pions are
present. This is not far worse, though, than in the case of set A. When the
pions are excluded the fit turns out to be very good in the case of set B
($\chi^2/dof=0.47$ and $0.0675$, respectively), proving these cases to be
completely compatible with a primordial quark-gluon phase.

In the case of $Pb+Pb$ the imposition of set B severely worsens the
quality of the fit. The situation cannot be remedied with the exclusion of
pions and $\chi^2/dof$ remains at the value of $18$. Also the fitted
temperature in case of set B is unrealistically high.

The findings concerning the $S+S$ and $S+Ag$ data are in agreement 
with the proximity of the chemical freeze out points of these experiments to \
the Statistical Bootstrap critical line that was found in [8]. On the
contrary the freeze out point of $Pb+Pb$ was not found to posses such an
attribute in [29], also in agreement with the present results.

In the case of RHIC the imposition of set B does not lead to a good
quality fit in the presence of pions ($\chi^2/dof=3.86$), but the fit turns
out to be quite good when the pions are excluded ($\chi^2/dof=1.2$), so the
thermodynamic parameters are compatible with a QGP phase.

The extracted parameters in case of set B are inserted to eqs.~(1) and (2)
and the parameters $V_0$ and $B^{1/4}$ are also determined. It is interesting
that in the cases of $S+S$, $S+Ag$ and $Au+Au$ (without the pions), which
have been proven compatible with set B, all the calculated values of $V_0$
and $B^{1/4}$ are close, compatible with a unique value for these parameters.

All the previous fits have been performed without the presence of the
pentaquark states. Similar fits have also been performed with the inclusion
of the pentaquarks. The difference in the extracted parameters, apart from
the volume, is found to be at most $1.4\%$ and in the parameter $VT^3$ at
most $6\%$ and so they are not listed.

The necessity of the expansion of the fugacity sector with the partial
equilibrium fugacities is also revealed with the application of the
present technique. If these fugacities are set to
$\gamma_u=\gamma_d=\gamma_s=1$ then the sector of the phase space that is
compatible with the QGP-hadron transition is severely limited.
In that case if a similar fit to the set B is performed, apart from the fact
that eqs.~(3) and (14) cannot be accommodated, the fit turns out to be worse.
The result in the case without the pions is then $\chi^2/dof=0.61$, $1.05$,
$26.1$ and $1.86$ for $S+S$, $S+Ag$, $Pb+Bb$ and $Au+Au$ respectively. In the
case of $Au+Au$ the compatibility with the QGP phase turns out to be dubious
now.

\vspace{0.3cm}
{\large \bf 6. Conclusions}

Although two different partition functions are used for the description
of the quark and hadronic side of matter, it is possible to preserve the
continuity of all chemical potentials and, of course, temperature at the
transition between one another, which is confined on a curve. Also, all the
constraints imposed by the
conservation laws of quantum quantities and the continuity of pressure can be
applied, leading, at the same time to a non-trivial solution of the
thermodynamic variables into a three quark flavour system. The key issue for
the success of this project is the expansion of the fugacity sector of the
available variables and the, already, introduced relative chemical
equilibrium variables can be used to serve that purpose.

The restrictions on the freeze-out conditions imposed by the existence of
a quark-gluon state in the early stages after a collision experiment can
be applied to every case that the thermalisation of the produced hadrons
has been proven. They can serve to separate the experiments compatible with
QGP state from those that are not.
In a simplified and quick to use form these restrictions acquire
the form of eqs.~(11) or (13).
The expansion of the fugacity sector with
the partial equilibrium fugacities, though, magnifies the part of the
phase space allowed by such constraints.

The whole methodology that was presented can be used for every grand
canonical partition function
adopted for the description of the HG or QGP phase. The inclusion of
interaction is crucial for the prediction of the critical point and the
volume expansion ratio, which could not be determined by the models used in
this work. At the moment lattice calculations have led to the determination
of the accurate quark-gluon equation of state with three quark flavours at
finite baryon chemical potential [20,30]. It would be interesting, though, if
these calculations could be extended with the inclusion of the
relative chemical equilibrium variables for light and strange quarks,
allowing matching with the existing hadron gas models. For the hadronic side
of matter the inclusion of the attractive part of interaction can be incorporated via
the statistical bootstrap [7,8], where the prediction of a critical point
is also possible [31]. The incorporation of the full set of parameters
$\gamma_i$ to these studies would allow for a more precise matching with a
primordial quark phase.

{\bf Acknowledgement} I would like to thank N. G. Antoniou, C. N. Ktorides
and F. K. Diakonos for fruitful discussions.

\begin{center}                                                                                                         
\begin{tabular}{|cc|cc|cc|cc|}
\hline
\multicolumn{2}{|c|}{$S+S$}           & \multicolumn{2}{|c|}{$S+Ag$}         & \multicolumn{2}{|c|}{$Pb+Pb$}         & \multicolumn{2}{|c|}{$Au+Au$}                 \\ \hline \hline
$K^+$                & $12.5\pm 0.4$  & ${K_s}^0$            & $15.5\pm 1.5$ & $N_p$                & $362\pm 5.1$   & $\Lambda$                  & $17.20\pm 1.75$  \\ 
$K^-$                & $6.9\pm 0.4$   & $\Lambda$            & $15.2\pm 1.2$ & $K^+$                & $103\pm 7.1$   & $\overline{\Lambda}$       & $12.15\pm 1.25$  \\
${K_s}^0$            & $10.5\pm 1.7$  & $\overline{\Lambda}$ & $2.6\pm 0.3$  & $K^-$                & $51.9\pm 3.6$  & $\Xi^-$                    & $2.11 \pm 0.23$  \\  
$\Lambda$            & $9.4\pm 1.0$   & $\overline{p}$       & $2.0\pm 0.8$  & ${K_s}^0$            & $81\pm 4$      & $\overline{\Xi}^+$         & $1.77 \pm 0.19$  \\ 
$\overline{\Lambda}$ & $2.2\pm 0.4$   & $p-\overline{p}$     & $43\pm 3$     & $\phi$               & $7.6\pm 1.1$   & $\Omega+\overline{\Omega}$ & $0.585\pm 0.150$ \\ 
$\overline{p}$       & $1.15\pm 0.40$ & $B-\overline{B}$     & $105\pm 12$   & $\Lambda$            & $53\pm 5$      & $p$                        & $26.37\pm 2.60$  \\ 
$p-\overline{p}$     & $21.2\pm 1.3$  & $h^{-(*)}$           & $186\pm 11$   & $\overline{\Lambda}$ & $4.64\pm 0.32$ & $\overline{p}$             & $18.72\pm 1.90$  \\
$B-\overline{B}$     & $54\pm 3$      &                      &               & $\Xi^-$              & $4.45\pm 0.22$ & ${K_s}^0$                  & $36.7 \pm 5.5$   \\ 
$h^{-(*)}$           & $98\pm 3$      &                      &               & $\overline{\Xi}^+$   & $0.83\pm 0.04$ & $\phi$                     & $5.73 \pm 0.78$  \\  
                     &                &                      &               & $\Omega$             & $0.62\pm 0.09$ & $K^{*0}$                   & $10.0 \pm 2.70$  \\  
                     &                &                      &               & $\overline{\Omega}$  & $0.20\pm 0.03$ & $\pi^{+(*)}$               & $239  \pm 10.6$  \\  
                     &                &                      &               & $\pi^{+(*)}$         & $619\pm 35.4$  & $\pi^{-(*)}$               & $239  \pm 10.6$  \\ 
                     &                &                      &               & $\pi^{-(*)}$         & $639\pm 35.4$  & $K^+/K^-$                  & $1.092\pm 0.023$ \\
                     &                &                      &               &                      &                & $\overline{K}^{*0}/K^{*0}$ & $0.92 \pm 0.27$  \\
                     &                &                      &               &                      &                & $\overline{\Omega}/\Omega$ & $0.95 \pm 0.16$  \\ \hline
\end{tabular}

{\footnotesize
$^{(*)}$ This multiplicity has not been used in the fits where the pions 
are excluded.} 
\end{center} 
 
\begin{center} 
Table 1. The full phase space multiplicities from the collision
experiments $S+S$ (NA35), $S+Ag$ (NA35) and $Pb+Pb$ (NA49), as well as
the midrapidity multiplicities and ratios from $Au+Au$ (STAR), used in the
fits. 
\end{center} 
\vspace{0.5cm}

\begin{center}
\begin{tabular}{|c|c|c|c|c|c|}
\hline
\multicolumn{2}{|c|}{}& $S+S$           & $S+Ag$          & $Pb+Pb$           & $Au+Au$         \\ \hline \hline
\multicolumn{6}{|c|}{fit with all}                                                            \\ \hline
set A&$\chi^2/dof$    & $4.03/3$        & $3.50/1$        & $16.4/7$          & $6.01/9$       \\ \hline
     &$\chi^2/dof$    & $14.7/5$        & $5.72/3$        & $162/9$           & $42.4/11$       \\
     &$T\;(MeV)$      & $243\pm16$      & $275\pm36$      & $436.5\pm7.5$     & $345\pm32$      \\
     &$\lambda_u$     & $1.536\pm0.022$ & $1.613\pm0.033$ & $1.668\pm0.020$   & $1.082\pm0.011$ \\
     &$\lambda_d$     & $1.534\pm0.022$ & $1.638\pm0.035$ & $1.728\pm0.022$   & $1.086\pm0.011$ \\
set B&$\gamma_u$      & $0.584\pm0.076$ & $0.458\pm0.091$ & $0.2730\pm0.0044$ & $0.349\pm0.033$ \\
     &$\gamma_d$      & $0.586\pm0.077$ & $0.472\pm0.093$ & $0.2937\pm0.0049$ & $0.381\pm0.035$ \\
     &$\gamma_s$      & $0.402\pm0.059$ & $0.309\pm0.063$ & $0.1880\pm0.0026$ & $0.322\pm0.031$ \\
     &$VT^3$          & $152.0\pm5.9$   & $279.6\pm8.5$   & $679\pm18$        & $377\pm26$      \\
     &$V_0^{(*)}$     & $5.4\pm1.2$     & $4.87\pm0.96$   & $2.332\pm0.083$   & $3.57\pm0.32$   \\
     &$B^{1/4}\;(MeV)$& $324\pm23$      & $358\pm49$      & $547.8\pm9.5$     & $442\pm42$      \\ \hline \hline
\multicolumn{6}{|c|}{fit without pions}                                                       \\ \hline
set A&$\chi^2/dof$    & $0.356/2$       & $0/0$           & $8.97/5$          & $2.12/7$        \\ \hline
     &$\chi^2/dof$    & $1.88/4$        & $0.135/2$       & $128/7$           & $10.8/9$        \\
     &$T\;(MeV)$      & $194.9\pm6.5$   & $209\pm19$      & $443.6\pm8.8$     & $221\pm24$      \\
     &$\lambda_u$     & $1.605\pm0.037$ & $1.661\pm0.043$ & $1.746\pm0.017$   & $1.075\pm0.011$ \\
     &$\lambda_d$     & $1.599\pm0.036$ & $1.695\pm0.046$ & $1.817\pm0.019$   & $1.081\pm0.011$ \\
set B&$\gamma_u$      & $0.949\pm0.095$ & $0.77\pm0.18$   & $0.2612\pm0.0042$ & $0.69\pm0.18$   \\
     &$\gamma_d$      & $0.958\pm0.097$ & $0.79\pm 0.18$  & $0.2801\pm0.0046$ & $0.74\pm0.19$   \\
     &$\gamma_s$      & $0.85\pm0.10$   & $0.60\pm0.16$   & $0.1987\pm0.0029$ & $0.79\pm0.23$   \\
     &$VT^3$          & $94.1\pm9.5$    & $199\pm40$      & $621\pm 16$       & $272\pm 53$     \\
     &$V_0^{(*)}$     & $6.8\pm1.4$     & $6.2\pm3.1$     & $2.216\pm0.093$   & $5.4\pm3.9$     \\
     &$B^{1/4}\;(MeV)$& $279\pm11$      & $291\pm29$      & $556\pm11$        & $305\pm37$      \\ \hline \hline
\end{tabular}
\end{center}

{\footnotesize
$^{(*)}$ $V_0$ is measured in $\;(10^{-11}\;MeV^{-4})$.}

\begin{center} 
Table 2. The results of fits on the $S+S$ (NA35), $S+Ag$ (NA35), $Pb+Pb$
(NA49) and $Au+Au$ (STAR) data with the imposition of the set of
constraints A and B, without the inclusion of the pentaquark states. 
\end{center}

{\bf Figure Captions}
\newtheorem{f}{Fig.} 
\begin{f} 
\rm Temperature as function of the baryon chemical potential at the
QGP-Hadron gas transition line, without and with the inclusion of the
pentaquark states.
\end{f}
\begin{f} 
\rm Relative chemical equilibrium variable of $u$-quark, $\gamma_u$, as
function of the baryon chemical potential at the QGP-Hadron gas transition
line, without and with the inclusion of the pentaquark states.
\end{f}
\begin{f} 
\rm Relative chemical equilibrium variable of $d$-quark, $\gamma_d$, as
function of relative chemical equilibrium variable of $u$-quark, $\gamma_u$,
at the QGP-Hadron gas transition line, without and with the inclusion of the
pentaquark states for the isospin symmetric case. The line
$\gamma_d=\gamma_u$ is also depicted.
\end{f}
\begin{f}
\rm Relative chemical equilibrium variable of $s$-quark, $\gamma_s$, as
function of the baryon chemical potential at the QGP-Hadron gas transition
line, without and with the inclusion of the pentaquark states.
\end{f}

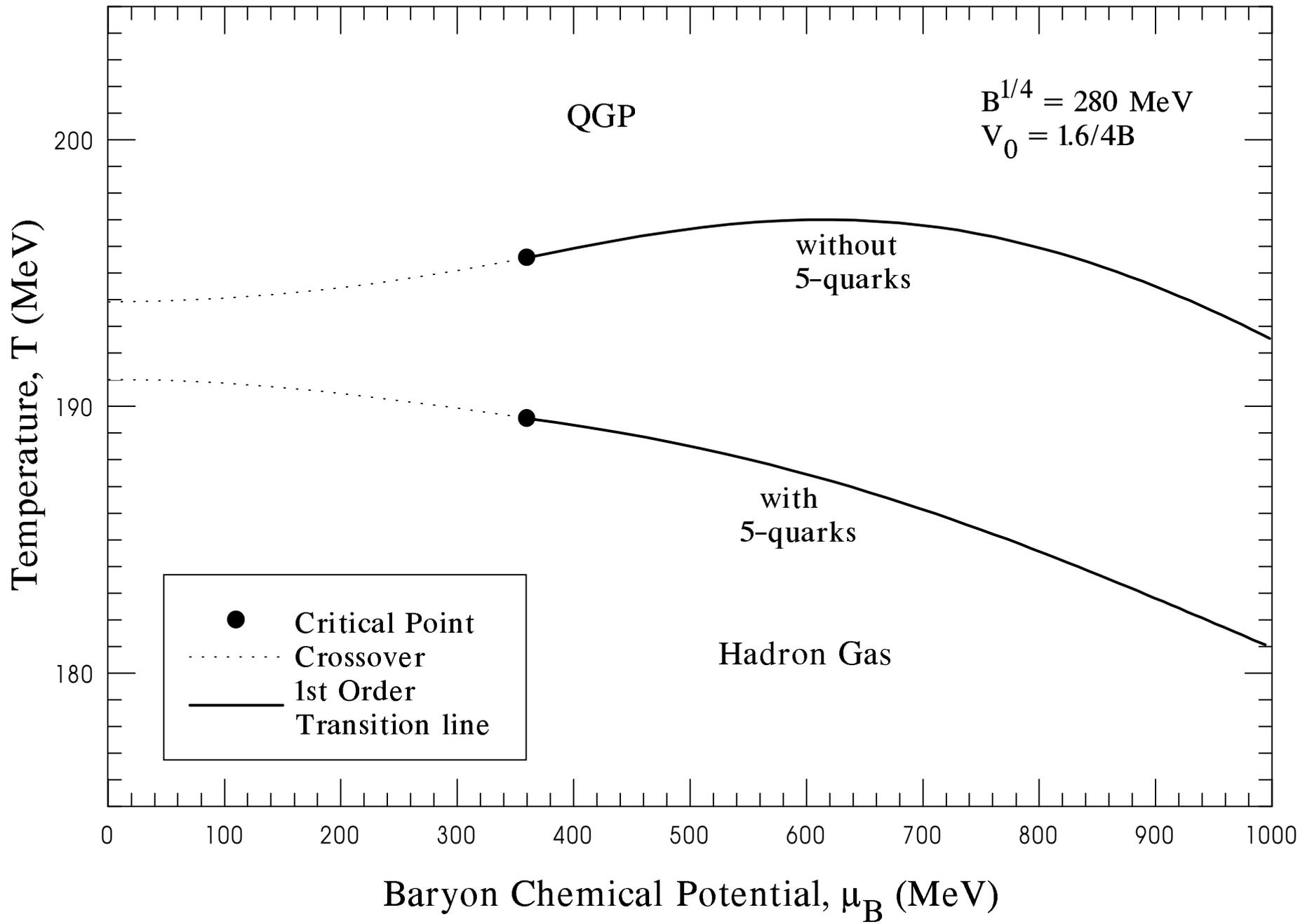
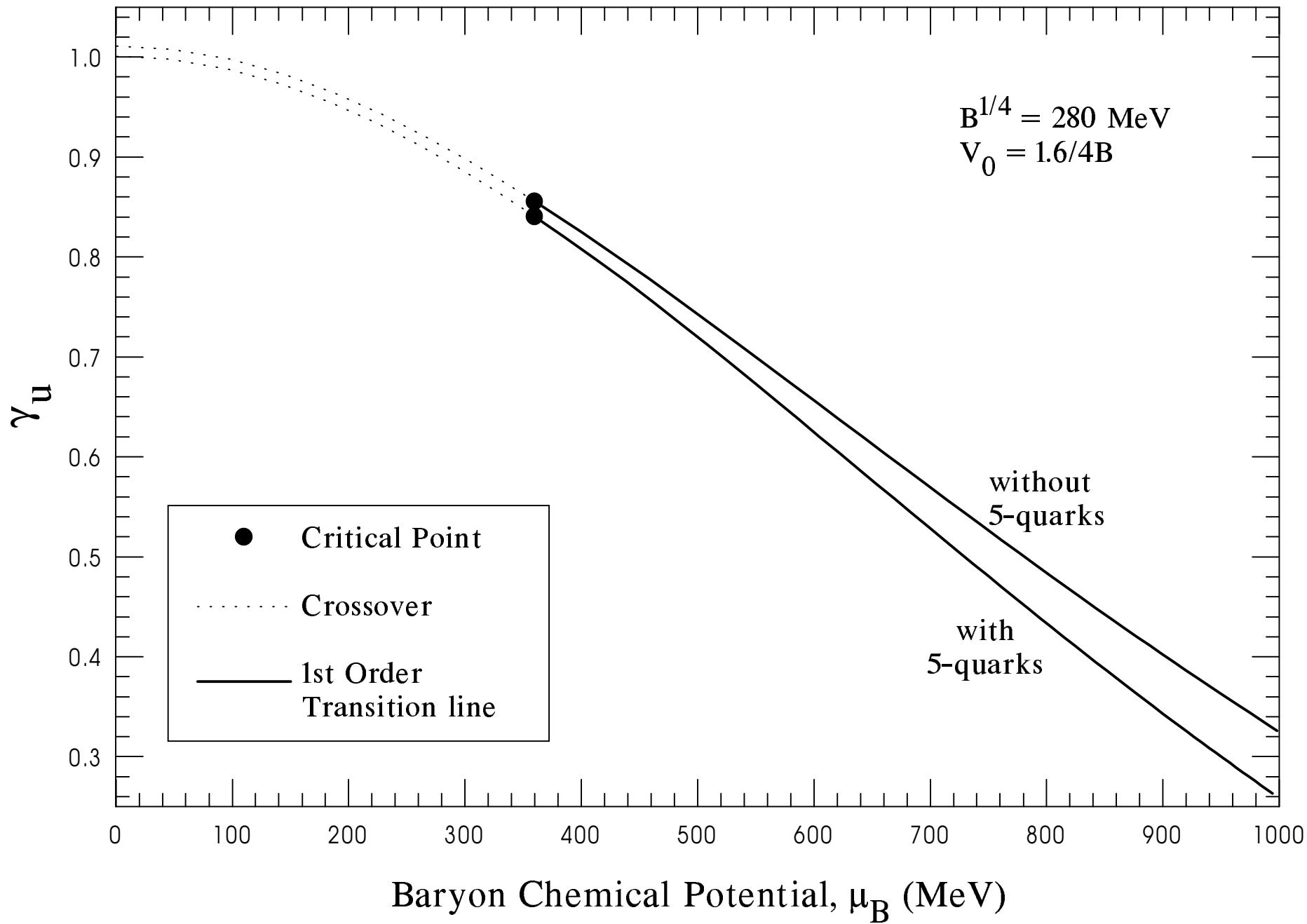
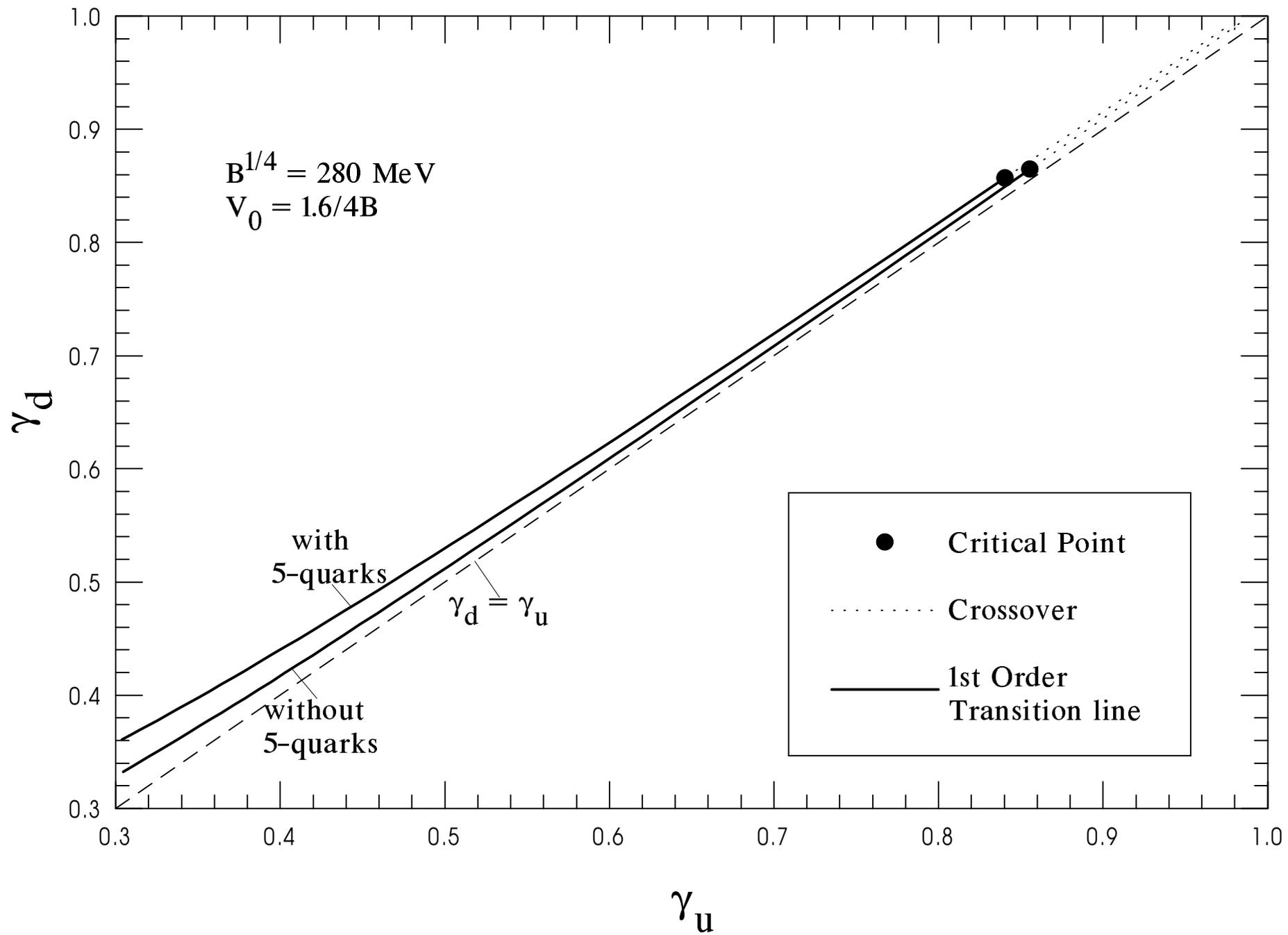
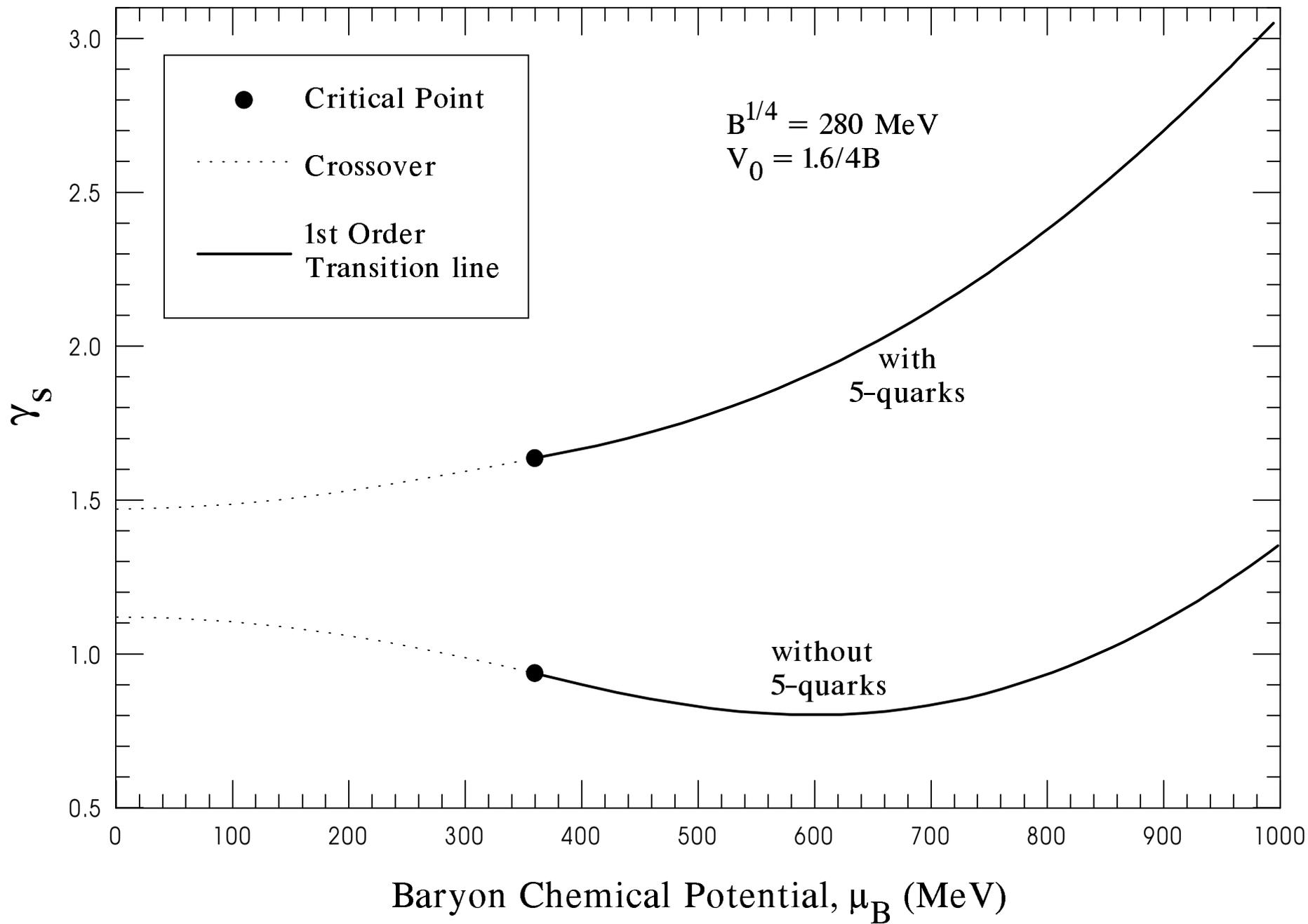
\end{document}